\documentclass[prd,aps,floats,preprint,preprintnumbers,amssymb,nofootinbib,longbibliography]{revtex4-1}
\usepackage{graphicx}
\usepackage{enumitem}
\usepackage{latexsym}
\usepackage{amsfonts}
\usepackage{amssymb}
\usepackage[dvipsnames]{xcolor}
\usepackage{CJK}
\usepackage[export]{adjustbox}
\usepackage{amsmath}
\usepackage{slashed}
\usepackage{dcolumn}
\usepackage{verbatim}
\usepackage{float}
\usepackage{multirow}
\usepackage{soul}
\usepackage[normalem]{ulem}
\usepackage{ulem}
\usepackage{hyperref}
\usepackage{adjustbox}
\usepackage{natbib}
\usepackage{epsfig}

\def\beq{\begin{equation}}
\def\eeq{\end{equation}}
\def\be{\begin{equation}}
\def\ee{\end{equation}}
\def\bea{\begin{eqnarray}}
\def\eea{\end{eqnarray}}

\def\r{\rho}

\def\m{\mu}
\def\n{\nu}

\def\r{\rho}
\def\s{\sigma}

\def\a{\alpha}
\def\b{\beta}
\def\d{\partial}
\def\g{\gamma}

\begin{document}

\title{ Kleinian black holes}
\author{Damien A. Easson\footnote{easson@asu.edu}}
\affiliation{
Department of Physics, Arizona State University, Tempe, AZ 85287-1504, USA}
\author{Max W. Pezzelle\footnote{max\_pezzelle@brown.edu}}
\affiliation{Brown Theoretical Physics Center, Department of Physics, Brown University, Providence, RI 02912, USA}

\date{\today}

\begin{abstract}
We prove there is a unique vacuum solution in split-signature spacetimes with Kleinian $SO(2,1)$ spherical symmetry.  We extend our analysis to accommodate a positive or negative cosmological constant and we prove the Kleinian spherically symmetric solutions to Einstein's equation are locally isomorphic to the split-signature analogues of Schwarzschild-(Anti)-de Sitter or Nariai spacetimes. Our analysis provides a Kleinian extension of Birkhoff's theorem to metrics with split-signature. Axisymmetric vacuum solutions are also considered, including $(2,2)$ signature formulations of the Kerr and Taub-NUT metrics.
\end{abstract}

\maketitle

\tableofcontents

\section{Introduction}
The Birkhoff theorem stands as a cornerstone in the realm of general relativity, affirming that any spherically symmetric vacuum solution derived from Einstein's equations can be locally equated to a corresponding segment of the Schwarzschild spacetime. The ubiquitous use of the theorem, along with its multiple origin history  are a testament to its profound importance in contemporary general relativity and cosmology, and naturally invite contemplation regarding the theorem's potential extension to diverse scenarios within the realm of gravitational physics and theories of modified gravity. 

Birkhoff's theorem states that the Schwarzschild metric:
\begin{equation}\label{schmet}
     \mathrm{d}s^2 =- \Big(1-\frac{2 m}{r}\Big)\mathrm{d}t^2 + \Big(1-\frac{2 m}{r}\Big)^{-1}\mathrm{d}r^2 + r^2\mathrm{d}\theta^2 + r^2\sin^2\theta\mathrm{d}\phi^2 \,,
\end{equation}
is the unique vacuum solution with spherical symmetry to the Einstein equations and there are no time-dependent solutions of this form, as long as $t$ remains the timelike coordinate and $r, \theta, \phi$ remain spacelike. The maximal analytic extension of the Schwarzschild solution has a non-static interior with spacelike (rather than timelike) Killing vector field. The theorem generalizes in the case of a non-vanishing cosmological constant $\Lambda$, to the Schwarzschild-(A)dS solutions. The caveat concerning the coordinate $t$ remaining timelike now extends to the region outside the de Sitter cosmological horizon which is time-dependent. The theorem is ubiquitously attributed to Birkhoff~\cite{Birkhoff1923}, although it was discovered earlier by Jebsen~\cite{Jebsen1921} and allegedly independently by Alexandrow~\cite{Alexandrow1923} and Eiesland~\cite{Eiesland1921,Eiesland1925}, in various forms (see e.g.~\cite{Deser:2004gi,Schleich:2009uj} for historical accounts).

The analytic continuation from Lorentzian $(1,3)$ signature to Kleinian $(2,2)$ split-signature has proved useful in the study of quantum field theory and quantum gravity. For example, as a tool to explore properties of physics in Lorentzian signature $\mathbb{L}^{3,1}$ \cite{Heckman:2022peq}. At the level of empty space, this translates into the analytic continuation from Minkowski space, $\mathbb{M}^{3,1}$ to Klein space, $\mathbb{K}^{2,2}$. 
Recently, studies of black holes written in Kleinian signature have appeared in the literature, making a complete investigation of their emergence from the vacuum theory desirable.  For example, it has been shown that linearized black hole geometries are captured by (2,2) three-point scattering amplitudes of a graviton and a massive spinning particle. The solutions include Kerr-Taub-NUT spacetimes which naturally encompass the Schwarzschild solution~\cite{Arkani-Hamed:2019ymq,Crawley:2021auj}. 

In addition, two-dimensional quantum states associated to four-dimensional linearized rotating self-dual black holes in split-signature have been shown to be comprised of global conformal primaries circulating on the celestial torus, the Kleinian analog of the celestial sphere. This allows for a direct connection to the S-matrix~\cite{Barrett:1993yn,Crawley:2023brz}. 

In some cases, it was shown that a spin parameter $a$ can be added to existing solutions via a simple coordinate change, including Self-dual Taub-NUT (SDTN) \cite{Crawley:2021auj} and the self-dual analog of Kerr~\cite{Easson:2023dbk}. 
Furthermore, a double copy relation between the gluon amplitude on a self-dual dyon and graviton amplitude on a SDTN spacetime was discussed in \cite{Adamo:2023fbj}. And a possible explanation of hidden symmetries of rotating black holes via analytic continuation to Kleinian signature was explored in \cite{Guevara:2023wlr}.

Given these advances in Kleinian geometry it is therefore natural to consider complete studies of black hole spacetimes with split signature, including properties of their uniqueness. In this paper we initiate this effort, and show that the Kleinian analog of Schwarzschild is the unique vacuum solution to the Einstein equations with Kleinian spherical symmetry and there are no time dependent solutions of this form, thus generalizing the Birkhoff theorem to split signature spacetimes. Spacetimes with a cosmological constant are also considered and we show the unique solutions are split-signature versions of Schwarzschild-(A)dS or Nariai Klein spaces. Finally, expanding our tolerance to allow for axially symmetric vacuum solutions, we conjure the Kleinian analogue of the Kerr solution applying the Newman-Janis algorithm to split-signature Schwarzschild written in Kerr-Schild coordinates, and discuss the Kleinian Taub-NUT solution.

This paper is organized as follows. In section~\ref{kleinspace}, we introduce flat Klein space and our conventions used to transition from Lorentzian to split signature Kleinian metrics. In section~\ref{vacsol}, we study Kleinian vacuum solutions to the Einstein equations, uncovering the Kleinian Schwarzschild solution.
In a brief discussion of black hole horizon crossing we find the Kleinian Kantowski-Sachs metric.
We then consider Schwarzschild-(Anti)-de Sitter and Nariai generalizations in the presence of non-vanishing cosmological constant. We then produce the Kleinian Schwarzschild solution in Kerr-Schild coordinates and using the Newman-Janis trick we construct a $(2,2)$ signature Kerr solution. In this family of metrics with axial symmetry we also discuss $(2,2)$ signature Taub-NUT space. We conclude in section~\ref{concl}. Appendix~A is included to demonstrate alternative analytic continuations from Lorentzian to Klein space.

\section{Klein Space}\label{kleinspace}
Let us begin with local Cartesian coordinates $x^\alpha = (t,x,y,z)$, in which the Lorentzian flat spacetime metric is
\begin{equation}\label{flatm}
    \mathrm{d}s^2 = - \mathrm{d}t^2 +\mathrm{d}x^2  + \mathrm{d}y^2 + \mathrm{d}z^2 \,.
\end{equation}
Transforming to spherical coordinates, the euclidean line element $ \mathrm{d}s^2 = g_{ij} \mathrm{d}x^i  \mathrm{d}x^j$ is then fixed by invariance of the metric on the manifold $\mathcal{M}$ under the $SO(3)$ rotation group: $SO(3) \times \mathcal{M} \rightarrow \mathcal{M}$ and we find
\be
\mathrm{d}s^2 = - \mathrm{d}t^2  + \mathrm{d}r^2 + r^2 \mathrm{d}\theta^2 + r^2  \sin^2 \theta \mathrm{d}\phi^2 \,.
\ee
This coordinate transformation is given by 
\begin{equation}\label{spherec}
    x = r\cos\phi\sin\theta \,, \qquad  y = r\sin\phi\sin\theta \,, \qquad  z = r\cos\theta \,,
\end{equation}
with inverse transformation
\begin{equation}\label{spherecinv}
r = \sqrt{x^2 + y^2 + z^2} \,, \qquad 
\theta = \arccos\left(\frac{z}{r}\right) \,, \qquad
\varphi  = \arctan\left(\frac{y}{x}\right)
\end{equation}
and the most general spherically symmetric spacetime metric is 
\begin{equation}\label{metgen}
    \mathrm{d}s^2 = - N(t,r)\mathrm{d}t^2  + P(t,r)\mathrm{d}r^2 + Q(r,t) \mathrm{d} \Omega_L^2\,,
\end{equation}
where $\mathrm{d} \Omega_L^2 = \mathrm{d}\theta^2 +  \sin^2 \theta \mathrm{d}\phi^2 $ is the metric on the unit sphere $\mathbf{S}^2$ and $N$, $P$, $Q$ are arbitrary functions of the time and radial coordinate.

It is possible to move from Lorentzian signature metrics to split-signature (Kleinian) metrics via analytic continuation. 
We may transition to Klein space in multiple ways.~\footnote{An alternative choice of continuation to the one presented here is provided in Appendix~A.} We choose to move to split signature metric via the analytic continuation
\begin{equation}\label{ttoit}
    t \rightarrow it \,, \qquad x \rightarrow ix\,, \qquad y \rightarrow iy \,,
\end{equation}
so that in Kleinian signature the flat metric (\ref{flatm}) becomes
\begin{equation}\label{xtoix}
    \mathrm{d}s^2 = \mathrm{d}t^2  -\mathrm{d}x^2 - \mathrm{d}y^2 +\mathrm{d}z^2 \,.
\end{equation}
The analytic continuation (\ref{ttoit}) is equivalent to the spherical coordinate euclideanization $\theta \rightarrow i \theta$ in (\ref{spherec}), so that $\cos{\theta} \rightarrow \cosh{\theta}$ and $\sin{\theta} \rightarrow i \sinh{\theta}$.
We can transform into coordinates where $SO(2,1)$ invariance is manifest via
\begin{equation}\label{stroct}
     x = r\cos\phi\sinh\theta \,, \qquad \qquad
     y = r\sin\phi\sinh\theta \,, \qquad \qquad
     z = r\cosh\theta \,.
\end{equation}
The inverse coordinate transformation is
\begin{equation}
    r = \sqrt{z^2 -x^2-y^2} \,, \qquad \theta = \mathrm{arccosh} \left(\frac{z}{\sqrt{z^2 -x^2-y^2}}\right)  \,, \qquad \phi = \arctan \left(\frac{y}{x}\right).
\end{equation}
The metric (\ref{xtoix}) becomes
\begin{equation}\label{kleinrthetaphi}
    \mathrm{d}s^2 = \mathrm{d}t^2 + \mathrm{d}r^2 - r^2\mathrm{d}\theta^2 - r^2\sinh^2\theta\mathrm{d}\phi^2.
\end{equation}
Invariance under the action of the Kleinian rotation symmetry group on the spacetime manifold yields corresponding infinitesimal motion generated by the complex Killing vector fields
\begin{eqnarray}
     \xi_x &=&   \sin{\phi}\, \d_\theta +  \coth{\theta} \cos{\phi} \,\d_ \phi \nonumber \\
     \xi_y &=& - \cos{\phi}\, \d_\theta +  \coth{\theta} \sin{\phi} \,\d_\phi \nonumber  \\
     \xi_z &=& \d_\phi 
\end{eqnarray}

Invariance under Kleinian rotations imply the vanishing of the Lie derivative $L_\xi g_{ij} = 0$ along these vector fields. 
The commutators of the above $[ \xi_i, \xi_j ] = \epsilon_{ijk} \xi_k $ form the Lie algebra of the Kleinian rotation group $SO(2,1)$, 
and thus we have invariance of the metric under the action $\xi: SO(2,1) \times \mathcal{M} \rightarrow \mathcal{M} $, of the group $SO(2,1)$ on the spacetime manifold $\mathcal{M}$. This suggests the most general Kleinian spherically symmetric metric may be written
\begin{equation}
    \mathrm{d}s^2 = N(T,R)\mathrm{d}T^2 + P(T,R)\mathrm{d}R^2 - Q(T,R)(\mathrm{d}\theta^2 + \sinh^2\theta\mathrm{d}\phi^2) \,,
\end{equation}
where $N$, $P$, and $Q$ are scalar functions of $T$ and $R$ only. We will also abbreviate the angular term in parentheses as $\mathrm{d}\theta^2 + \sinh^2\theta\mathrm{d}\phi^2 \equiv \mathrm{d}\Omega^2$, the unit radius hyperbolic space with constant Ricci scalar curvature $R = -2$. We can make a diffeomorphism to coordinates where $Q = r^2$ for some coordinate $r$: this transforms the metric into
\begin{equation}
    \mathrm{d}s^2 = \Bigg(N + P\Big(\frac{\partial R}{\partial T}\Big)^2\Bigg)\mathrm{d}T^2 + P\frac{\partial R}{\partial T}\frac{\partial R}{\partial r}(\mathrm{d}T\mathrm{d}r + \mathrm{d}r\mathrm{d}T) + P\Big(\frac{\partial R}{\partial r}\Big)^2\mathrm{d}r^2 - r^2\mathrm{d}\Omega^2.
\end{equation}
Defining new metric profile functions $g_{TT}$, $g_{Tr}$, and $g_{rr}$, we see that a Kleinian spherically symmetric metric can also be written as
\begin{equation}\label{metricTrcoords}
    \mathrm{d}s^2 = g_{TT}\mathrm{d}T^2 + g_{Tr}(\mathrm{d}T\mathrm{d}r + \mathrm{d}r\mathrm{d}T) + g_{rr}\mathrm{d}r^2 - r^2\mathrm{d}\Omega^2.
\end{equation}
Next, we show that this metric can be written in the form
\begin{equation}\label{so21metric}
    \mathrm{d}s^2 = n(t,r)\mathrm{d}t^2 + p(t,r)\mathrm{d}r^2 - r^2\mathrm{d}\Omega^2.
\end{equation}
For if we make the coordinate transformation $\mathrm{d}t = \frac{\partial t}{\partial T}\mathrm{d}T + \frac{\partial t}{\partial r}\mathrm{d}r$ on the above line element, we have that
\begin{equation}
    \mathrm{d}s^2 = n\Big(\frac{\partial t}{\partial T}\Big)^2\mathrm{d}T^2 + n\frac{\partial t}{\partial T}\frac{\partial t}{\partial r}(\mathrm{d}T\mathrm{d}r + \mathrm{d}r\mathrm{d}T) + \Bigg(n\Big(\frac{\partial t}{\partial r}\Big)^2 + p\Bigg)\mathrm{d}r^2 - r^2\mathrm{d}\Omega^2.
\end{equation}
Matching the above line element to \eqref{metricTrcoords}, we obtain
\begin{equation}
    n\Big(\frac{\partial t}{\partial T}\Big)^2 = g_{TT}\mathrm{,\ } n\frac{\partial t}{\partial T}\frac{\partial t}{\partial r} = g_{Tr}\mathrm{,\ and\ } n\Big(\frac{\partial t}{\partial r}\Big)^2 + p = g_{rr}.
\end{equation}
These are three differential equations for three unknown functions $t(T,r)$, $n(t,r)$, and $p(t,r)$, which is enough to specify each along with supplying initial conditions on $t$. The most general Kleinian spherically symmetric metric is given by \eqref{so21metric}.

\section{Kleinian vacuum solution}\label{vacsol}
We now begin an exploration of Kleinian solutions to the vacuum Einstein equations $G_{\mu\nu} = 0$, where as usual $G_{\mu\nu}$ is the Einstein tensor defined in terms of the Ricci tensor and scalar,
$G_{\mu\nu} = R_{\mu\nu} - \frac{1}{2} g_{\mu\nu} R$.

\subsection{Kleinian Schwarzschild}
Here we produce the slit-signature generalization of the Birkhoff theorem, by searching for spherically symmetric solutions to the vaccum Einstein equations. We substitute (\ref{so21metric}) into the vacuum Einstein equations $G_{\mu\nu} = 0$ to find that
\begin{equation}
    G_{tr} = \frac{\partial_t p}{pr} = 0 \,,
\end{equation}
which implies that $p$ is a function independent of time. With this condition, no time derivatives of $n$ appear in the other equations, and this implies that $n$ is independent of time as well. $G_{tt} = 0$ implies that
\begin{equation}
    p(p-1) + r\partial_r p = 0,
\end{equation}
which has the solution $p = (1+A/r)^{-1}$ with $A$ a constant. Additionally $G_{rr} = 0$ implies that
\begin{equation}
    1-p+\frac{r\partial_r n}{n} = 0.
\end{equation}
Substituting the solution for $p$ into the above equation allows us to solve for $n$: we find that $n = B/A + B/r$ with $B$ a constant. If we impose that the metric reduce to its flat form (\ref{kleinrthetaphi}) as $r\rightarrow\infty$ then $A=B$. We discover that the unique spherically symmetric solution to the vacuum Einstein equations (\ref{so21metric}) is given by
\begin{equation}\label{kschw}
    \mathrm{d}s^2 = \Big(1+\frac{A}{r}\Big)\mathrm{d}t^2 + \Big(1+\frac{A}{r}\Big)^{-1}\mathrm{d}r^2 - r^2\mathrm{d}\theta^2 - r^2\sinh^2\theta\mathrm{d}\phi^2.
\end{equation}
This is the Kleinian form of the familiar Schwarzschild metric, as identified through analytic continuation $t \rightarrow it $, $\theta \rightarrow i \theta$ and by setting integration constant $A = -2 m$, returning us to (\ref{schmet}). Note that in the derivation one has the choice of  analytic continuation to Klein space and as to whether the orbit of SO(2,1) is spacelike (as we have done here), giving the two-dimensional hyperboloid $\mathbb{H}^{2}$, or timelike (as in Appendix A, see Eq.~(\ref{sch2}), giving two-dimensional de Sitter spacetime $(\mathrm{d}\mathbb{S})_2$.
\subsubsection{Kleinian Kantowski-Sachs}
Here we consider moving across $r=2m$ in the Kleinien Schwarzschild black hole solution.
In Lorentzian Schwarzschild spacetime the metric function $g_{tt} = (1 - 2 m/r)$ changes sign while passing to the black hole interior $r< 2m$.
This is related to the changing of the roles of space and time as seen by the transition of $t$ and $r$ to become the spacelike and timelike coordinates respectively. The interior Schwarzschild metric is not static but rather a homogeneous and anisotropic Kantowski-Sachs metric
\begin{equation}
    \mathrm{d}s^2 = -\nu(t)^{-1}\mathrm{d}t^2 + \nu(t)\mathrm{d}r^2 + h^2(t)(\mathrm{d}\theta^2 + \sin^2\theta\mathrm{d}\phi^2)\,,
\end{equation}
with two scale factors, $\nu(t) = (A/t-1)$ and $h(t) = t$. Under the analytic continuation leading to (\ref{kschw}), we see that crossing $r=2m = -A$
leads to an Euclidean metric with signature $(-,-,-,-)$! This is in stark contrast to what happens in ordinary Lorentzian Schwarzschild when crossing the horizon. We note that this transition to a Euclidean metric does not occur if one chooses to analytically continue Lorentzian Schwarzschild to Klein space via the transformation leading to (\ref{sch2}). 

In \cite{Crawley:2021auj}, split-signature linearized metrics were constructed from purely on-shell radiative modes. Under the analytic continuation used above, the Kleinian traced-reverse linearized metric has a real integrand, and ultimately reduces to a convergent Gaussian integral (in the region of interest). In contrast, under the continuation leading to (\ref{sch2}), the integrand for the metric is imaginary and oscillates, requiring an $i \epsilon$ prescription for convergence. 
We leave further physical interpretation of this phenomenon to future work.
\subsection{(A)dS Generalization}
If we allow for a nonzero cosmological constant, then the Einstein equations state $G_{\mu\nu} + \Lambda g_{\mu\nu} = 0$. We proceed as in the previous section and substitute the Kleinian spherically symmetric metric \eqref{so21metric} into the field equations. The equation $G_{tr} = -\Lambda g_{tr}$ implies that $\partial_t p = 0$, as before; this further implies that $\partial_t n = 0$ from the other field equations. The equation $G_{tt} = -\Lambda g_{tt}$ demands that
\begin{equation}
    p(p-1) + r\partial_r p = \Lambda pr^2.
\end{equation}
This has the solution $p = (1+A/r-\Lambda r^2/3)^{-1}$. Then the $G_{rr}= -\Lambda g_{rr}$ field equations imply that
\begin{equation}
    1-p+\frac{r\partial_r n}{n} = -\Lambda pr^2.
\end{equation}
After enforcing that $n\rightarrow 1$ as $r\rightarrow\infty$ as before, this has the solution $n = 1+A/r-\Lambda r^2/3 = p^{-1}$. The unique solution with metric asatz $\ref{so21metric})$ and nonzero cosmological constant is then
\begin{equation}
    \mathrm{d}s^2 = \Big(1+\frac{A}{r} - \frac{\Lambda}{3}r^2\Big)\mathrm{d}t^2 + \Big(1+\frac{A}{r} - \frac{\Lambda}{3}r^2\Big)^{-1}\mathrm{d}r^2 - r^2\mathrm{d}\theta^2 - r^2\sinh^2\theta\mathrm{d}\phi^2 \,,
\end{equation}
which is the Kleinian Schwarzshild-(A)dS solution corresponding to asymptotically Anti-de Sitter/de Sitter space for $\Lambda<0$ or $\Lambda>0$, respectively. 

\subsection{Kleinian Nariai}
There is one further possibility to consider. Returning to our general metric (\ref{metgen}) in Kleinian signature
\begin{equation}\label{metgen2}
    \mathrm{d}s^2 =  N(t,r)\mathrm{d}t^2  + P(t,r)\mathrm{d}r^2 - Q(r,t) \mathrm{d} \Omega^2\,,
\end{equation}
where again $ \mathrm{d} \Omega^2 = \mathrm{d} \theta^2 + \sinh^2{\theta}\mathrm{d}\phi^2$. The Einstein equation for this metric along with vacuum energy support,  $G_{\mu\nu} + \Lambda g_{\mu\nu}=0$, gives
\begin{equation}\label{lvacrt}
    \frac{{\dot{P}} Q'}{{2  P Q}} + \frac{{ N' \dot{Q}}}{{2  N Q}} + \frac{{Q' \dot{Q}}}{{2  Q^2}} - \frac{{\dot{Q}'}}{{Q}}=0 \,,
\end{equation}
\begin{equation}\label{lvactt}
 -\frac{N}{Q} - \frac{N P'  Q' }{2 P^2 Q} - \frac{N Q'^2}{4 P Q^2} + \frac{N  Q''}{P Q} + \frac{\dot{P} \dot Q}{2 P Q} + \frac{\dot{Q}^2}{4 Q^2} = - N \Lambda \,,
\end{equation}
\begin{equation}
  -\frac{P}{Q} + \frac{N'  Q'}{2 N Q} + \frac{ Q'^2}{4 Q^2} - \frac{P \dot N \dot Q}{2 N^2 Q} - \frac{P \dot Q^2}{4 N Q^2} + \frac{P \ddot Q}{N Q}= -P \Lambda \,.
\end{equation}

The aforementioned case of interest occurs when the Kleinian sphere orbits of  the $SO(2,1)$ isometry have the same constant radius of $r= \sqrt{Q}$. A detailed discussion may be found in \cite{Morrow-Jones:1993mwp}. In this case from (\ref{lvactt}) with constant $Q$ we find
\begin{equation}
    Q = \frac{1}{\Lambda}\,.
\end{equation}
We further discover
\begin{equation}
    N = P^{-1}= 1 - \Lambda r^2 \,,
\end{equation}
which gives the Kleinian analogue of the Nariai spacetime
\begin{equation}
    \mathrm{d}s^2 = \Big(1 - \Lambda r^2\Big)\mathrm{d}t^2 + \Big(1 - \Lambda r^2 \Big)^{-1}\mathrm{d}r^2 - \Lambda^{-1} (\mathrm{d}\theta^2 +\sinh^2\theta\mathrm{d}\phi^2)\,,
\end{equation}
whose corresponding Lorentzian metric analogue is only well defined for $\Lambda>0$.

\subsection{(2,2) signature Kerr}
We now discuss the case of nonzero angular momentum with spin parameter $a$. We will arrive at our solution using arguments analogous to the Newman-Janis algorithm \cite{Newman:1965tw} and it is instructive to reformulate our Kleinian Schwarzschild Birkhoff argument of section \ref{vacsol} using Kerr-Schild coordinates. 
\subsubsection{Kleinian Kerr-Schild coordinates}\label{kerrschildcoord}
The metric (\ref{so21metric}) may be put into Kerr-Schild form \cite{Bini:2010hrs}

\begin{equation}\label{ksg}
    g_{\mu\nu}= \eta_{\mu\nu} + \Phi k_\mu k_\nu \,,
\end{equation}
where $k^\sigma$ is a null vector and for our purposes $\Phi$ is a function of $r$.
For this discussion we will move to split signature with the continuation: $\theta \rightarrow i \theta$ and $\phi \rightarrow i \phi$, which allows for a real Kerr-Schild null vector and a real metric. This is equivalent to analytically continuing $z \rightarrow iz$ in Cartesian coordinates leading to flat space metric (\ref{flatz}).~\footnote{An alternative choice of continuation leading to a complex Kerr-Schild null vector and complex metric is presented in Appendix~A.}  We will use the coordinate transformation convention (\ref{stroct}).
Taking the null vector
\begin{equation}
     k^\sigma = (1,1,0,0)
\end{equation}
we see $k_\mu k^\mu = 0$ with respect to both the flat metric 
\begin{equation}
    \mathrm{d}s^2 = -\mathrm{d}t^2 + \mathrm{d}r^2 - r^2\mathrm{d}\theta^2 +r^2\sinh^2\theta\mathrm{d}\phi^2 \,,
\end{equation}
and  the full Kleinian Kerr-Schild metric (\ref{ksg}), having line element
\begin{equation}
   ds^2 = - \mathrm{d}t^2 + \mathrm{d}r^2  - r^2 \mathrm{d}\theta^2 + r^2  \sinh^2\theta \mathrm{d}\phi^2+ (\mathrm{d}r - \mathrm{d}t)^2 \,\Phi(r) \,.
\end{equation}
Note the appearance of cross-terms $drdt$ in the metric remains compatible with spherical $SO(3)$ symmetry. 
With this ansatz we compute the Ricci scalar $R$ and set it to zero as is appropriate for a vacuum solution
\begin{equation}\label{rscal}
   R =  \Phi''(r) + \frac{2 (\Phi(r) + 2r\Phi'(r))}{r^2} = 0 \,,
\end{equation}
where prime denotes differentiation with respect to $r$.
We find the solution for $\Phi$ is
\begin{equation}
    \Phi(r) = \frac{c_1}{r^2} + \frac{c_2}{r} \,.
\end{equation}
Further demanding the vacuum condition on the Ricci scalar $R_{\mu\nu}=0$ yields $c_1 = 0$.
Hence setting $c_2 = A$ we find the Schwarzschild solution written in Kerr-Schild form with 
\begin{equation}\label{phir}
    \Phi(r) = \frac{A}{r} \,,
\end{equation}
and the Kleinian Schwarzschild metric written in Kerr-Schild coordinates is thus
\begin{equation}\label{kksm}
 \mathrm{d}s^2 = - \left(1 - \frac{A}{r}\right) \mathrm{d}t^2+\left(1 + \frac{A}{r} \right)   \mathrm{d}r^2  - r^2  \mathrm{d}\theta^2 + r^2  \sinh^2\theta  \mathrm{d}\phi^2 - 2\frac{A }{r}  \mathrm{d}r  \mathrm{d}t\,.
\end{equation}
The curvature invariants computed from (\ref{kksm}) are identical to Eqs.~(\ref{rscals}).
The familiar Lorentzian Schwarzschild metric in Kerr-Schild form is recovered from the above by taking 
\begin{equation}
(\phi \rightarrow i \phi) \,, \qquad (\theta \rightarrow i \theta) \,,
\end{equation}
and is given by line element
\begin{equation}\label{fins}
     \mathrm{d}s^2 = - \left(1 - \frac{A}{r}\right)  \mathrm{d}t^2 + \left(1 +\frac{A}{r} \right)  \mathrm{d}r^2  + r^2  \mathrm{d}\theta^2 + r^2  \sin^2\theta  \mathrm{d}\phi^2 - 2 \frac{A}{r}  \mathrm{d}r  \mathrm{d}t \,.
\end{equation}

\subsubsection{ Kleinian Kerr solution}\label{kkerr}
Having discovered Kleinian Schwarzschild in Kerr-Schild coordiantes, we may now formulate the Kerr solution in $(2,2)$ signature, by employing the Newmann-Janis trick \cite{Newman:1965tw} compexifying $z \rightarrow z - i a$, so that the Kerr-Schild scalar for Kleinian Schwarzschild becomes
\begin{equation}
    \Phi = \frac{A}{\sqrt{-x^2 - y^2 + (z - ia)^2}} \,.
\end{equation}
Introducing oblate spheroidal coordinates
\begin{equation}
    x + i y = (r + ia) \sinh{\theta} e^{i \phi}\,, \qquad z = r \cosh{\theta} \,,
\end{equation}
the generating Kerr-Schild scalar is given by
\begin{equation}
    \Phi = \frac{ A}{r +i a \cosh{\theta} } \,.
\end{equation}
We define
\begin{eqnarray}
    \alpha &=& Re \Phi = \frac{A}{2} \left( \frac{1}{r + i a \cosh{\theta}} +  \frac{A}{r -i a \cosh{\theta}} \right) = \frac{r}{r^2 + a^2 \cosh^2{\theta}} \,, \\
     \beta &=& Im \Phi = \frac{A}{2 i} \left( \frac{1}{r +i a \cosh{\theta}} -  \frac{A}{r -i a \cosh{\theta}} \right) = \frac{- a \cosh{\theta}}{r^2 + a^2 \cosh^2{\theta} } \,,
\end{eqnarray}
which yields the Kerr space-time in Cartesian Kerr–Schild form with
\begin{equation}
     A \alpha = \frac{2 m r}{r^2 + a^2 \cosh^2{\theta}} = \frac{2 m r^3}{r^4 + a^2 z^2} \,,
\end{equation}
where $A = 2m$ and Kerr-Schild metric
\begin{equation}
    g_{\mu\nu} = \eta_{\mu\nu} + \frac{2 m r^3}{r^4 + a^2 z^2} k_\mu k_\nu
\end{equation}
with null vector in terms of the coordinates introduced in section \ref{kleinspace},
\begin{equation}
    k_\sigma = i \left( 1, \frac{ xr + a y}{r^2 + a^2},\frac{ yr - a x}{r^2 + a^2}, \frac{- iz}{r} \right) \,.
\end{equation}
This concludes our derivation of the Kleinian Kerr solution. 

\subsection{(2,2) signature Taub-NUT}
As a second example of a stationary axisymmetric spacetime we consider Lorentzian Taub-NUT \cite{Taub:1950ez,Newman:1963yy} given by
\begin{equation}
     \mathrm{d}s^2 = - f(r)(\mathrm{d}t - 2 N \cos{\theta} \mathrm{d}\phi)^2 + \frac{ \mathrm{d}r^2}{f(r)} + (r^2 + N^2)(\mathrm{d}\theta^2 + \sin^2{\theta} \mathrm{d}\phi^2 )\,,
\end{equation}
where 
\begin{equation}
    f(r) = \frac{r^2 - 2 m r - N^2}{r^2 + N^2} \,.
\end{equation}
In the above $m$ is the ADM mass and $N$ is the so-called NUT charge. We may transition to Klein space via the complex rotations (\ref{tthetac}) along with $N \rightarrow iN$ to arrive at the Kleinian Taub-NUT line element
\begin{equation}
    \mathrm{d}s^2 =  f(r)(\mathrm{d}t - 2 N \cosh{\theta} \mathrm{d}\phi)^2 + \frac{\mathrm{d}r^2}{f(r)} - (r^2 - N^2)(\mathrm{d}\theta^2 + \sinh^2{\theta} \mathrm{d}\phi^2 )\,.
\end{equation}
For an extensive analysis of this space see \cite{Crawley:2021auj}. Let us focus on the (anti)-self-dual case where $m = \pm N$, so that 
\begin{equation}
    f(r) = \frac{r + m}{r - m} \,.
\end{equation}
The metric for the self-dual case $m = +N$ becomes
\begin{equation}\label{sdtnone}
    \mathrm{d}s^2 =  \frac{r - m}{r+ m}(\mathrm{d}t - 2 m \cosh{\theta} \mathrm{d}\phi)^2 + \frac{r + m}{r - m} \mathrm{d}r^2 - (r^2 - m^2)(\mathrm{d}\theta^2 + \sinh^2{\theta} \mathrm{d}\phi^2 )\,.
\end{equation}
Interestingly, taking $r \rightarrow r - m$,  puts the SDTN metric (\ref{sdtnone}) in the form
\begin{eqnarray}\label{sdtnb}
    \mathrm{d}s^2 &=& \Big(1-\frac{2m}{r}\Big)\mathrm{d}t^2 + \Big(1-\frac{2m}{r}\Big)^{-1}\mathrm{d}r^2 - r^2\mathrm{d}\theta^2 - r^2\sinh^2\theta\mathrm{d}\phi^2 \nonumber \\
    &-& 4m \Big(1-\frac{2m}{r}\Big) \cosh \theta \, \mathrm{d}t \, \mathrm{d}\phi + 2m r \,\mathrm{d}\theta^2\nonumber \\ 
    &+&  2m \Big(2m\Big(1-\frac{2m}{r}\Big) \cosh^2 \theta +  r \sinh^2 \theta\Big) \mathrm{d} \phi^2 \,.
\end{eqnarray}
Glancing back at (\ref{kschw}), we see that the SDTN may be written as ordinary Kleinian Schwarzschild (identifying $A = -2m$) plus additional terms given by the second and third lines of (\ref{sdtnb}). The additional terms are made of the cross-term $g_{t \phi}$ and an extra $g_{\theta \theta}$ and $g_{\phi \phi}$ piece, breaking the Kleinian spherical symmetry of (\ref{kschw}). Remarkably, the curvature invariants for (\ref{sdtnb}) are identical to (\ref{rscals}), and unlike the Kerr metric, depend only on the $r$ coordinate.
Of course, the above metric is Ricci flat $R_{\mu\nu}=R=0$. All components of the Riemann tensor and curvature invariants go to zero as $r \rightarrow \infty$.

As a final comment we note that unlike their electromagnetic linear counterparts, Lorentzian (real) gravitational solutions are not in one-to-one correspondence with self-dual solutions. However, for Kerr-Schild spacetimes there is a simple correspondence between real vacuum solutions and self-dual spacetimes due to the Einstein equations linearizing for Kerr-Schild metrics. This fact was emphasized in \cite{Easson:2023dbk}, which also presented a self-dual version of the Kerr solution.

\section{Conclusions}\label{concl}
We have completed a study of the uniqueness of Kleinian black hole solutions with maximally symmetric vacua. A comprehensive study of Kleinian vacuum solutions is essential, given that  split-signature metrics and their associated properties have emerged as a rich and prominent subject of study in contemporary physics. In this paper we have demonstrated that there is a uniqueness theorem for $SO(2,1)$ symmetric metrics in Kleinian signature analogous to Birkhoff's theorem in Lorentzian signature. 

In vacuum, the unique Kleinian-spherically symmetric metric is a complexified version of the Schwarzschild metric. In a spacetime with nonzero cosmological constant, the unique solution is generalized to a complexified version of Schwarzschild-(A)dS or Nariai space. Additionally, we have written down the unique metric form in Kerr-Schild coordinates, and through the Newman-Janis transform we were able to construct the Kleinian Kerr metric. Finally we considered Kleinian Taub-NUT showing that it can be written as Kleinian Schwarzschild with additional terms, further enriching our understanding of axisymmetric solutions in split-signature. A natural continuation of this work would be to extend the analysis to Kleinian electrovac solutions and we leave such exploration to future research.

\section*{Acknowledgements}
We are delighted to thank Lars Aalsma, Samarth Chawla, Gabe Herczeg, Cindy Keeler, Tucker Manton, Maulik Parikh and Andy Strominger for useful discussions and correspondence.  
DAE is supported in part by the U.S. Department of Energy, Office of High Energy Physics, under Award Number DE-SC0019470.
\newpage
\section*{Appendix}\label{append}
\subsection{Alternative analytic continuation}
An alternative choice of continuation from Lorentzian metric (\ref{flatm}) to Kleinian space is the minimal complexification on $z \rightarrow iz$ alone, so that flat Klein space is given by metric
\begin{equation}\label{flatz}
    \mathrm{d}s^2 = -\mathrm{d}t^2 -\mathrm{d}z^2 + \mathrm{d}x^2 + \mathrm{d}y^2.
\end{equation}
Transforming to spherical coordinates with
\begin{equation}
    x = r\cos\phi\cosh\theta \,, \qquad y = r\sin\phi\cosh\theta \,, \qquad z = r\sinh\theta\,.
\end{equation}
The inverse coordinate transformation is
\begin{equation}
    r = \sqrt{x^2+y^2-z^2} \mathrm{,\ } \phi = \tan^{-1}\frac{y}{x} \mathrm{,\ } \theta = \sinh^{-1}\frac{z}{\sqrt{x^2+y^2-z^2}}.
\end{equation}
The flat space metric is then
\begin{equation}\label{kleinrthetaphii}
    \mathrm{d}s^2 = -\mathrm{d}t^2 + \mathrm{d}r^2 - r^2\mathrm{d}\theta^2 + r^2\cosh^2\theta\mathrm{d}\phi^2.
\end{equation}
Under this continuation the Kleinian Schwarzschild metric is
\begin{equation}\label{sch2}
    \mathrm{d}s^2 = -\Big(1+\frac{A}{r}\Big)\mathrm{d}t^2 + \Big(1+\frac{A}{r}\Big)^{-1}\mathrm{d}r^2 - r^2\mathrm{d}\theta^2 + r^2\cosh^2\theta\mathrm{d}\phi^2.
\end{equation}
Naturally, the choice of transformation does not matter for the spacetime structure and corresponding curvature invariants for 
either analytic continuation prescription are equivalent. Some quadratic, cubic and quartic curvature invariants are found to be
\begin{eqnarray}
R_{\m\n\s\r}R^{\m\n\s\r} &=& \frac{12A^2}{r^6} \,, \nonumber \\
R_{\m\n}{}^{\a\b}R_{\a\b}{}^{\r\s}R_{\r\s}{}^{\m\n} &=& -\frac{12A^3}{r^9}\,, \nonumber \\
\nabla_\a R_{\b\g\s\r} \nabla^\a R^{\b\g\s\r} &=& \frac{180 A^2 (A+r)}{r^9}\,,  \nonumber \\
R_{\alpha\beta\gamma\rho} R^{\alpha\beta\gamma}{}_{\sigma} R_{\mu\nu\kappa}{}^{\rho} R^{\mu\nu\kappa\sigma} &=& \frac{36 A^4}{r^{12}}\,. \label{rscals}
\end{eqnarray}

As another example of alternative continuations, the arguments leading to the Kleinian Schwarzschild solution in Kerr-Schild coordinates of section \ref{kerrschildcoord} are easily repeated using an alternative analytic continuation. One may
journey to Klein space via the continuation $t \rightarrow i t$, $\theta \rightarrow i \theta$, thusly accommodating  a complex null vector:
\begin{equation}
     k^\sigma = (i,1,0,0)
\end{equation}
so that $k_\mu k^\mu = 0$ with respect to both the flat metric (\ref{kleinrthetaphi}) and the full metric (\ref{kschwcomp}). 
Our Kleinian Kerr-Schild line element from (\ref{ksg}) is now the complex metric
\begin{equation}\label{kschwcomp}
    \mathrm{d}s^2 =  \mathrm{d}t^2+  \mathrm{d}r^2   - r^2  \mathrm{d}\theta^2 - r^2  \sinh^2\theta  \mathrm{d}\phi^2 + ( \mathrm{d}r + i\, \mathrm{d}t)^2\,\Phi(r)
\end{equation}
The Kleinian Schwarzschild metric written in Kerr-Schild form under the above continuation reads
\begin{equation}\label{kksmcomp}
 \mathrm{d}s^2 = \left(1 - \frac{A}{r}\right) \mathrm{d}t^2+\left(1 + \frac{A}{r} \right)   \mathrm{d}r^2  - r^2  \mathrm{d}\theta^2 - r^2  \sinh^2\theta  \mathrm{d}\phi^2 + \frac{2iA  \mathrm{d}r  \mathrm{d}t}{r} \,.
\end{equation}
The curvature invariants computed from (\ref{kksmcomp}), Eqs.~(\ref{rscals}) remain identical.
The real Schwarzschild result (\ref{fins}) is recovered from the above by taking 
\begin{equation}\label{tthetac}
(t \rightarrow i t) \,, \qquad (\theta \rightarrow i \theta) \,.
\end{equation}

\bibliography{bib}

\begin{thebibliography}{20}%
\makeatletter
\providecommand \@ifxundefined [1]{%
 \@ifx{#1\undefined}
}%
\providecommand \@ifnum [1]{%
 \ifnum #1\expandafter \@firstoftwo
 \else \expandafter \@secondoftwo
 \fi
}%
\providecommand \@ifx [1]{%
 \ifx #1\expandafter \@firstoftwo
 \else \expandafter \@secondoftwo
 \fi
}%
\providecommand \natexlab [1]{#1}%
\providecommand \enquote  [1]{``#1''}%
\providecommand \bibnamefont  [1]{#1}%
\providecommand \bibfnamefont [1]{#1}%
\providecommand \citenamefont [1]{#1}%
\providecommand \href@noop [0]{\@secondoftwo}%
\providecommand \href [0]{\begingroup \@sanitize@url \@href}%
\providecommand \@href[1]{\@@startlink{#1}\@@href}%
\providecommand \@@href[1]{\endgroup#1\@@endlink}%
\providecommand \@sanitize@url [0]{\catcode `\\12\catcode `\$12\catcode `\&12\catcode `\#12\catcode `\^12\catcode `\_12\catcode `\%12\relax}%
\providecommand \@@startlink[1]{}%
\providecommand \@@endlink[0]{}%
\providecommand \url  [0]{\begingroup\@sanitize@url \@url }%
\providecommand \@url [1]{\endgroup\@href {#1}{\urlprefix }}%
\providecommand \urlprefix  [0]{URL }%
\providecommand \Eprint [0]{\href }%
\providecommand \doibase [0]{http://dx.doi.org/}%
\providecommand \selectlanguage [0]{\@gobble}%
\providecommand \bibinfo  [0]{\@secondoftwo}%
\providecommand \bibfield  [0]{\@secondoftwo}%
\providecommand \translation [1]{[#1]}%
\providecommand \BibitemOpen [0]{}%
\providecommand \bibitemStop [0]{}%
\providecommand \bibitemNoStop [0]{.\EOS\space}%
\providecommand \EOS [0]{\spacefactor3000\relax}%
\providecommand \BibitemShut  [1]{\csname bibitem#1\endcsname}%
\let\auto@bib@innerbib\@empty
\bibitem [{\citenamefont {Birkhoff}(1923)}]{Birkhoff1923}%
  \BibitemOpen
  \bibfield  {author} {\bibinfo {author} {\bibfnamefont {G.~D.}\ \bibnamefont {Birkhoff}},\ }\href@noop {} {\bibfield  {journal} {\bibinfo  {journal} {Relativity and Modern Physics, Harvard University Press,}\ ,\ \bibinfo {pages} {p. 253}} (\bibinfo {year} {1923})}\BibitemShut {NoStop}%
\bibitem [{\citenamefont {Jebsen}(1921)}]{Jebsen1921}%
  \BibitemOpen
  \bibfield  {author} {\bibinfo {author} {\bibfnamefont {J.T.}\ \bibnamefont {Jebsen}},\ }\href@noop {} {\bibfield  {journal} {\bibinfo  {journal} {Ark. Mat. Ast. Fys.}\ }\textbf {\bibinfo {volume} {15}},\ \bibinfo {pages} {no. 18, 1} (\bibinfo {year} {1921})}\BibitemShut {NoStop}%
\bibitem [{\citenamefont {Alexandrow}(1923)}]{Alexandrow1923}%
  \BibitemOpen
  \bibfield  {author} {\bibinfo {author} {\bibfnamefont {W.}~\bibnamefont {Alexandrow}},\ }\href@noop {} {\bibfield  {journal} {\bibinfo  {journal} {Ann. Physik}\ }\textbf {\bibinfo {volume} {72}},\ \bibinfo {pages} {141} (\bibinfo {year} {1923})}\BibitemShut {NoStop}%
\bibitem [{\citenamefont {Eiesland}(1921)}]{Eiesland1921}%
  \BibitemOpen
  \bibfield  {author} {\bibinfo {author} {\bibfnamefont {J.}~\bibnamefont {Eiesland}},\ }\href@noop {} {\bibfield  {journal} {\bibinfo  {journal} {Amer. Math. Soc. Bull.}\ }\textbf {\bibinfo {volume} {27}},\ \bibinfo {pages} {410} (\bibinfo {year} {1921})}\BibitemShut {NoStop}%
\bibitem [{\citenamefont {Eiesland}(1925)}]{Eiesland1925}%
  \BibitemOpen
  \bibfield  {author} {\bibinfo {author} {\bibfnamefont {J.}~\bibnamefont {Eiesland}},\ }\href@noop {} {\bibfield  {journal} {\bibinfo  {journal} {Trans. Amer. Math. Soc.}\ }\textbf {\bibinfo {volume} {27}},\ \bibinfo {pages} {213} (\bibinfo {year} {1925})}\BibitemShut {NoStop}%
\bibitem [{\citenamefont {Deser}\ and\ \citenamefont {Franklin}(2005)}]{Deser:2004gi}%
  \BibitemOpen
  \bibfield  {author} {\bibinfo {author} {\bibfnamefont {Stanley}\ \bibnamefont {Deser}}\ and\ \bibinfo {author} {\bibfnamefont {J.}~\bibnamefont {Franklin}},\ }\bibfield  {title} {\enquote {\bibinfo {title} {{Schwarzschild and Birkhoff a la Weyl}},}\ }\href {\doibase 10.1119/1.1830505} {\bibfield  {journal} {\bibinfo  {journal} {Am. J. Phys.}\ }\textbf {\bibinfo {volume} {73}},\ \bibinfo {pages} {261--264} (\bibinfo {year} {2005})},\ \Eprint {http://arxiv.org/abs/gr-qc/0408067} {arXiv:gr-qc/0408067} \BibitemShut {NoStop}%
\bibitem [{\citenamefont {Schleich}\ and\ \citenamefont {Witt}(2010)}]{Schleich:2009uj}%
  \BibitemOpen
  \bibfield  {author} {\bibinfo {author} {\bibfnamefont {Kristin}\ \bibnamefont {Schleich}}\ and\ \bibinfo {author} {\bibfnamefont {Donald~M.}\ \bibnamefont {Witt}},\ }\bibfield  {title} {\enquote {\bibinfo {title} {{A simple proof of Birkhoff's theorem for cosmological constant}},}\ }\href {\doibase 10.1063/1.3503447} {\bibfield  {journal} {\bibinfo  {journal} {J. Math. Phys.}\ }\textbf {\bibinfo {volume} {51}},\ \bibinfo {pages} {112502} (\bibinfo {year} {2010})},\ \Eprint {http://arxiv.org/abs/0908.4110} {arXiv:0908.4110 [gr-qc]} \BibitemShut {NoStop}%
\bibitem [{\citenamefont {Heckman}\ \emph {et~al.}(2022)\citenamefont {Heckman}, \citenamefont {Joyce}, \citenamefont {Sakstein},\ and\ \citenamefont {Trodden}}]{Heckman:2022peq}%
  \BibitemOpen
  \bibfield  {author} {\bibinfo {author} {\bibfnamefont {Jonathan~J.}\ \bibnamefont {Heckman}}, \bibinfo {author} {\bibfnamefont {Austin}\ \bibnamefont {Joyce}}, \bibinfo {author} {\bibfnamefont {Jeremy}\ \bibnamefont {Sakstein}}, \ and\ \bibinfo {author} {\bibfnamefont {Mark}\ \bibnamefont {Trodden}},\ }\bibfield  {title} {\enquote {\bibinfo {title} {{Exploring 2~+~2 answers to 3~+~1 questions}},}\ }\href {\doibase 10.1142/S0217751X22502013} {\bibfield  {journal} {\bibinfo  {journal} {Int. J. Mod. Phys. A}\ }\textbf {\bibinfo {volume} {37}},\ \bibinfo {pages} {2250201} (\bibinfo {year} {2022})},\ \Eprint {http://arxiv.org/abs/2208.02267} {arXiv:2208.02267 [hep-th]} \BibitemShut {NoStop}%
\bibitem [{\citenamefont {Arkani-Hamed}\ \emph {et~al.}(2020)\citenamefont {Arkani-Hamed}, \citenamefont {Huang},\ and\ \citenamefont {O'Connell}}]{Arkani-Hamed:2019ymq}%
  \BibitemOpen
  \bibfield  {author} {\bibinfo {author} {\bibfnamefont {Nima}\ \bibnamefont {Arkani-Hamed}}, \bibinfo {author} {\bibfnamefont {Yu-tin}\ \bibnamefont {Huang}}, \ and\ \bibinfo {author} {\bibfnamefont {Donal}\ \bibnamefont {O'Connell}},\ }\bibfield  {title} {\enquote {\bibinfo {title} {{Kerr black holes as elementary particles}},}\ }\href {\doibase 10.1007/JHEP01(2020)046} {\bibfield  {journal} {\bibinfo  {journal} {JHEP}\ }\textbf {\bibinfo {volume} {01}},\ \bibinfo {pages} {046} (\bibinfo {year} {2020})},\ \Eprint {http://arxiv.org/abs/1906.10100} {arXiv:1906.10100 [hep-th]} \BibitemShut {NoStop}%
\bibitem [{\citenamefont {Crawley}\ \emph {et~al.}(2022)\citenamefont {Crawley}, \citenamefont {Guevara}, \citenamefont {Miller},\ and\ \citenamefont {Strominger}}]{Crawley:2021auj}%
  \BibitemOpen
  \bibfield  {author} {\bibinfo {author} {\bibfnamefont {Erin}\ \bibnamefont {Crawley}}, \bibinfo {author} {\bibfnamefont {Alfredo}\ \bibnamefont {Guevara}}, \bibinfo {author} {\bibfnamefont {Noah}\ \bibnamefont {Miller}}, \ and\ \bibinfo {author} {\bibfnamefont {Andrew}\ \bibnamefont {Strominger}},\ }\bibfield  {title} {\enquote {\bibinfo {title} {{Black holes in Klein space}},}\ }\href {\doibase 10.1007/JHEP10(2022)135} {\bibfield  {journal} {\bibinfo  {journal} {JHEP}\ }\textbf {\bibinfo {volume} {10}},\ \bibinfo {pages} {135} (\bibinfo {year} {2022})},\ \Eprint {http://arxiv.org/abs/2112.03954} {arXiv:2112.03954 [hep-th]} \BibitemShut {NoStop}%
\bibitem [{\citenamefont {Barrett}\ \emph {et~al.}(1994)\citenamefont {Barrett}, \citenamefont {Gibbons}, \citenamefont {Perry}, \citenamefont {Pope},\ and\ \citenamefont {Ruback}}]{Barrett:1993yn}%
  \BibitemOpen
  \bibfield  {author} {\bibinfo {author} {\bibfnamefont {John~W.}\ \bibnamefont {Barrett}}, \bibinfo {author} {\bibfnamefont {G.~W.}\ \bibnamefont {Gibbons}}, \bibinfo {author} {\bibfnamefont {M.~J.}\ \bibnamefont {Perry}}, \bibinfo {author} {\bibfnamefont {C.~N.}\ \bibnamefont {Pope}}, \ and\ \bibinfo {author} {\bibfnamefont {P.}~\bibnamefont {Ruback}},\ }\bibfield  {title} {\enquote {\bibinfo {title} {{Kleinian geometry and the N=2 superstring}},}\ }\href {\doibase 10.1142/S0217751X94000650} {\bibfield  {journal} {\bibinfo  {journal} {Int. J. Mod. Phys. A}\ }\textbf {\bibinfo {volume} {9}},\ \bibinfo {pages} {1457--1494} (\bibinfo {year} {1994})},\ \Eprint {http://arxiv.org/abs/hep-th/9302073} {arXiv:hep-th/9302073} \BibitemShut {NoStop}%
\bibitem [{\citenamefont {Crawley}\ \emph {et~al.}(2023)\citenamefont {Crawley}, \citenamefont {Guevara}, \citenamefont {Himwich},\ and\ \citenamefont {Strominger}}]{Crawley:2023brz}%
  \BibitemOpen
  \bibfield  {author} {\bibinfo {author} {\bibfnamefont {Erin}\ \bibnamefont {Crawley}}, \bibinfo {author} {\bibfnamefont {Alfredo}\ \bibnamefont {Guevara}}, \bibinfo {author} {\bibfnamefont {Elizabeth}\ \bibnamefont {Himwich}}, \ and\ \bibinfo {author} {\bibfnamefont {Andrew}\ \bibnamefont {Strominger}},\ }\bibfield  {title} {\enquote {\bibinfo {title} {{Self-dual black holes in celestial holography}},}\ }\href {\doibase 10.1007/JHEP09(2023)109} {\bibfield  {journal} {\bibinfo  {journal} {JHEP}\ }\textbf {\bibinfo {volume} {09}},\ \bibinfo {pages} {109} (\bibinfo {year} {2023})},\ \Eprint {http://arxiv.org/abs/2302.06661} {arXiv:2302.06661 [hep-th]} \BibitemShut {NoStop}%
\bibitem [{\citenamefont {Easson}\ \emph {et~al.}(2023)\citenamefont {Easson}, \citenamefont {Herczeg}, \citenamefont {Manton},\ and\ \citenamefont {Pezzelle}}]{Easson:2023dbk}%
  \BibitemOpen
  \bibfield  {author} {\bibinfo {author} {\bibfnamefont {Damien~A.}\ \bibnamefont {Easson}}, \bibinfo {author} {\bibfnamefont {Gabriel}\ \bibnamefont {Herczeg}}, \bibinfo {author} {\bibfnamefont {Tucker}\ \bibnamefont {Manton}}, \ and\ \bibinfo {author} {\bibfnamefont {Max}\ \bibnamefont {Pezzelle}},\ }\bibfield  {title} {\enquote {\bibinfo {title} {{Isometries and the double copy}},}\ }\href {\doibase 10.1007/JHEP09(2023)162} {\bibfield  {journal} {\bibinfo  {journal} {JHEP}\ }\textbf {\bibinfo {volume} {09}},\ \bibinfo {pages} {162} (\bibinfo {year} {2023})},\ \Eprint {http://arxiv.org/abs/2306.13687} {arXiv:2306.13687 [gr-qc]} \BibitemShut {NoStop}%
\bibitem [{\citenamefont {Adamo}\ \emph {et~al.}(2023)\citenamefont {Adamo}, \citenamefont {Bogna}, \citenamefont {Mason},\ and\ \citenamefont {Sharma}}]{Adamo:2023fbj}%
  \BibitemOpen
  \bibfield  {author} {\bibinfo {author} {\bibfnamefont {Tim}\ \bibnamefont {Adamo}}, \bibinfo {author} {\bibfnamefont {Giuseppe}\ \bibnamefont {Bogna}}, \bibinfo {author} {\bibfnamefont {Lionel}\ \bibnamefont {Mason}}, \ and\ \bibinfo {author} {\bibfnamefont {Atul}\ \bibnamefont {Sharma}},\ }\bibfield  {title} {\enquote {\bibinfo {title} {{Scattering on self-dual Taub-NUT}},}\ }\href@noop {} {\  (\bibinfo {year} {2023})},\ \Eprint {http://arxiv.org/abs/2309.03834} {arXiv:2309.03834 [hep-th]} \BibitemShut {NoStop}%
\bibitem [{\citenamefont {Guevara}\ and\ \citenamefont {Kol}(2023)}]{Guevara:2023wlr}%
  \BibitemOpen
  \bibfield  {author} {\bibinfo {author} {\bibfnamefont {Alfredo}\ \bibnamefont {Guevara}}\ and\ \bibinfo {author} {\bibfnamefont {Uri}\ \bibnamefont {Kol}},\ }\bibfield  {title} {\enquote {\bibinfo {title} {{Black Hole Hidden Symmetries from the Self-Dual Point}},}\ }\href@noop {} {\  (\bibinfo {year} {2023})},\ \Eprint {http://arxiv.org/abs/2311.07933} {arXiv:2311.07933 [hep-th]} \BibitemShut {NoStop}%
\bibitem [{\citenamefont {Morrow-Jones}\ and\ \citenamefont {Witt}(1993)}]{Morrow-Jones:1993mwp}%
  \BibitemOpen
  \bibfield  {author} {\bibinfo {author} {\bibfnamefont {J.}~\bibnamefont {Morrow-Jones}}\ and\ \bibinfo {author} {\bibfnamefont {D.~M.}\ \bibnamefont {Witt}},\ }\bibfield  {title} {\enquote {\bibinfo {title} {{Inflationary initial data for generic spatial topology}},}\ }\href {\doibase 10.1103/PhysRevD.48.2516} {\bibfield  {journal} {\bibinfo  {journal} {Phys. Rev. D}\ }\textbf {\bibinfo {volume} {48}},\ \bibinfo {pages} {2516--2528} (\bibinfo {year} {1993})}\BibitemShut {NoStop}%
\bibitem [{\citenamefont {Newman}\ and\ \citenamefont {Janis}(1965)}]{Newman:1965tw}%
  \BibitemOpen
  \bibfield  {author} {\bibinfo {author} {\bibfnamefont {E.~T.}\ \bibnamefont {Newman}}\ and\ \bibinfo {author} {\bibfnamefont {A.~I.}\ \bibnamefont {Janis}},\ }\bibfield  {title} {\enquote {\bibinfo {title} {{Note on the Kerr spinning particle metric}},}\ }\href {\doibase 10.1063/1.1704350} {\bibfield  {journal} {\bibinfo  {journal} {J. Math. Phys.}\ }\textbf {\bibinfo {volume} {6}},\ \bibinfo {pages} {915--917} (\bibinfo {year} {1965})}\BibitemShut {NoStop}%
\bibitem [{\citenamefont {Bini}\ \emph {et~al.}(2010)\citenamefont {Bini}, \citenamefont {Geralico},\ and\ \citenamefont {Kerr}}]{Bini:2010hrs}%
  \BibitemOpen
  \bibfield  {author} {\bibinfo {author} {\bibfnamefont {Donato}\ \bibnamefont {Bini}}, \bibinfo {author} {\bibfnamefont {Andrea}\ \bibnamefont {Geralico}}, \ and\ \bibinfo {author} {\bibfnamefont {Roy~P.}\ \bibnamefont {Kerr}},\ }\bibfield  {title} {\enquote {\bibinfo {title} {{The Kerr-Schild ansatz revised}},}\ }\href {\doibase 10.1142/S0219887810004518} {\bibfield  {journal} {\bibinfo  {journal} {Int. J. Geom. Meth. Mod. Phys.}\ }\textbf {\bibinfo {volume} {7}},\ \bibinfo {pages} {693} (\bibinfo {year} {2010})},\ \Eprint {http://arxiv.org/abs/1408.4601} {arXiv:1408.4601 [gr-qc]} \BibitemShut {NoStop}%
\bibitem [{\citenamefont {Taub}(1951)}]{Taub:1950ez}%
  \BibitemOpen
  \bibfield  {author} {\bibinfo {author} {\bibfnamefont {A.~H.}\ \bibnamefont {Taub}},\ }\bibfield  {title} {\enquote {\bibinfo {title} {{Empty space-times admitting a three parameter group of motions}},}\ }\href {\doibase 10.2307/1969567} {\bibfield  {journal} {\bibinfo  {journal} {Annals Math.}\ }\textbf {\bibinfo {volume} {53}},\ \bibinfo {pages} {472--490} (\bibinfo {year} {1951})}\BibitemShut {NoStop}%
\bibitem [{\citenamefont {Newman}\ \emph {et~al.}(1963)\citenamefont {Newman}, \citenamefont {Tamburino},\ and\ \citenamefont {Unti}}]{Newman:1963yy}%
  \BibitemOpen
  \bibfield  {author} {\bibinfo {author} {\bibfnamefont {E.}~\bibnamefont {Newman}}, \bibinfo {author} {\bibfnamefont {L.}~\bibnamefont {Tamburino}}, \ and\ \bibinfo {author} {\bibfnamefont {T.}~\bibnamefont {Unti}},\ }\bibfield  {title} {\enquote {\bibinfo {title} {{Empty space generalization of the Schwarzschild metric}},}\ }\href {\doibase 10.1063/1.1704018} {\bibfield  {journal} {\bibinfo  {journal} {J. Math. Phys.}\ }\textbf {\bibinfo {volume} {4}},\ \bibinfo {pages} {915} (\bibinfo {year} {1963})}\BibitemShut {NoStop}%
\end{thebibliography}%


\end{document}